\documentclass[10pt,preprint]{aastex}
\usepackage{amsmath,amssymb}

\def\bmath#1{{\bf #1}}

\usepackage[numberedappendix]{emulateapj5}
\usepackage{apjfonts}

\journalinfo{The Astrophysical Journal, \rm 629: scheduled for 2005 August 10 
{\rm [e-print~{\tt astro-ph/0501362}]}}

\def\exref#1{(\ref{#1})}

\def\eqref#1{Eq.~(\ref{#1})}
\def\eqsref#1{Eqs.~(\ref{#1})}

\def\Eqref#1{Equation~(\ref{#1})}
\def\Eqsref#1{Equations~(\ref{#1})}

\def\const{{\rm const}}

\def\bea{\begin{eqnarray}}
\def\eea{\end{eqnarray}}

\def\({\left(}
\def\){\right)}
\def\<{\langle}
\def\>{\rangle}
\def\lt{\left}
\def\rt{\right}
\def\bl{\bigl}
\def\br{\bigr}

\def\dd{\partial}
\def\diff{d}

\def\vdel{\bmath{\nabla}}
\def\delpar{\nabla_{\parallel}}

\def\vperp{v_\perp}
\def\vpar{v_\parallel}
\def\vu{\bmath{u}}

\def\vB{\bmath{B}}

\def\vb{\bmath{\hat b}}

\def\pperp{p_\perp}
\def\ppar{p_\parallel}
\def\Tperp{T_\perp}
\def\Tpar{T_\parallel}

\def\Brms{B_{\rm rms}}
\def\Bbend{B_{\rm bend}}

\def\rbend{\rho_{i,{\rm bend}}}

\def\vth{v_{\rm th}}

\def\uf{U}

\def\ld{l_{\nu}}
\def\lf{L}
\def\lres{l_{\eta}}
\def\mfp{\lambda_{\rm mfp}}

\def\nui{\nu_{ii}}

\def\nuB{\nu_{\rm B}}
\def\nuQL{\nu_{\rm QL}}

\def\vk{\bmath{k}}
\def\kpar{k_{\parallel}}
\def\kperp{k_{\perp}}
\def\vkperp{\vk_{\perp}}
\def\kd{k_{\nu}}

\def\dvu{\delta\vu} 
\def\dvuperp{\delta\vu_\perp} 
\def\dupar{\delta u_\parallel}

\def\dvB{\delta\vB}

\def\dB{\delta B}
\def\dvb{\delta\vb} 
\def\dpperp{\delta\pperp}
\def\dppar{\delta\ppar}

\def\vA{v_A}
\def\gmax{\gamma_{\rm max}}

\def\Re{{\rm Re}}
\def\Rm{{\rm Rm}}

\def\Pm{{\rm Pm}}
\def\ReQL{\Re_{\rm QL}}

\slugcomment{\today}
\shorttitle{PLASMA INSTABILITIES IN CLUSTERS OF GALAXIES} 
\shortauthors{SCHEKOCHIHIN ET AL.}

\begin{document}

\title{PLASMA INSTABILITIES AND MAGNETIC FIELD GROWTH IN CLUSTERS OF GALAXIES} 
\author{A.~A.~Schekochihin,\altaffilmark{1}
S.~C.~Cowley,\altaffilmark{2}
R.~M.~Kulsrud,\altaffilmark{3} 
G.~W.~Hammett,\altaffilmark{3}  
and P.~Sharma\altaffilmark{3}}
\altaffiltext{1}{Department of Applied Mathematics and 
Theoretical Physics, University of Cambridge, 
Wilberforce Road, Cambridge~CB3~0WA, UK; as629@damtp.cam.ac.uk.}
\altaffiltext{2}{Department of Physics and Astronomy, 
UCLA, Los Angeles, CA~90095-1547; and 
Plasma Physics Group, Imperial College London, 
Blackett Laboratory, Prince Consort Road, 
London~SW7~2BW, UK.}
\altaffiltext{3}{Plasma Physics Laboratory, Princeton University, 
P.~O.~Box 451, Princeton, NJ 08543.} 

\begin{abstract}
We show that under very general conditions, cluster plasmas 
threaded by weak magnetic fields are subject to very fast growing 
plasma instabilities driven by the anisotropy of the 
plasma pressure (viscous stress) with respect to the local direction 
of the magnetic field. Such an anisotropy will 
naturally arise in any weakly magnetized plasma that has low 
collisionality and is subject to stirring. 
The magnetic field must be sufficiently weak 
for the instabilities to occur, viz., $\beta\gtrsim\Re^{1/2}$. 
The instabilities are captured by the extended MHD model with Braginskii viscosity. 
However, their growth rates are proportional to the wavenumber 
down to the ion gyroscale, so MHD equations with Braginskii viscosity 
are not well posed and a fully kinetic treatment is necessary. 
The instabilities can lead to magnetic fields in clusters 
being amplified from seed strength of $\sim10^{-18}$~G to dynamically 
important strengths of $\sim10~\mu$G on cosmologically trivial time scales 
($\sim10^8$~yr). The fields produced during the amplification stage 
are at scales much smaller than observed. Predicting the saturated field scale 
and structure will require a kinetic theory of magnetized cluster turbulence. 
\end{abstract}


\section*{}
\vskip-0.5cm

Plasma in clusters of galaxies is likely 
to be in a turbulent state. 
Cluster mergers produce bulk flows at scales $\lf\sim 1$~Mpc 
with velocities $\uf\sim10^2...10^3$~km s$^{-1}$. 
Assuming an intracluster medium (ICM) with temperatures 
$T\sim10^7...10^8$~K, these bulk velocities are comparable 
to the thermal (or sound) speed~$\vth=(T/m_i)^{1/2}$ 
(here $m_i$ is the ion mass and, for simplicity, we take the 
ICM to be made of hydrogen plasma with equal electron and 
ion temperatures). Flows of this magnitude 
can drive turbulence in the ICM. 
The characteristic time for the turbulence to be 
established is $\lf/\uf\sim10^9$~yr. 
Numerical simulations of cluster evolution 
\citep{Norman_Bryan,Roettinger_Stone_Burns,Ricker_Sarazin} 
support this qualitative picture. 
In recent years there has been a rising interest in measuring 
the cluster turbulence both with future 
instruments \citep[{\em Astro}-E2; see][]{Inogamov_Sunyaev} and 
with present ones \citep[{\em XMM-Newton}; see][]{Schuecker_etal}.

The ICM is fully ionized. 
The ion kinematic viscosity is $\nu\sim\vth^2/\nui$, 
where $\nui=4\pi n e^4\ln\Lambda m_i^{-1/2}T^{-3/2}$ is the ion-ion 
collision frequency, $n\sim10^{-2}...10^{-3}$~cm$^{-3}$ is the ion 
number density (for cluster cores), $e$~is the electron charge, 
and $\ln\Lambda\sim20$ is the Coulomb logarithm 
\citep[e.g.,][]{Helander_Sigmar}. 
This gives Reynolds numbers in the range $\Re\sim10^2...10^3$. 
As the turbulent velocities at the outer scales are 
$\lesssim\vth$, the turbulence in the inertial range will be 
subsonic. It is natural to assume that Kolmogorov's dimensional 
theory should apply at least approximately. Then the 
viscous scale of the turbulence is $\ld\sim\lf\Re^{-3/4}\sim10...30$~kpc. 
These numbers appear to agree quite well with observations 
of turbulence in the Coma cluster via pressure maps \citep{Schuecker_etal}. 

If we estimate the magnetic diffusivity of the ICM 
using the standard \citet{Spitzer_book} formula 
$\eta\sim T^{-3/2}m_e^{1/2} e^2 c^2\ln\Lambda/4\pi$, 
we find an extremely small value leading 
to enormous magnetic Reynolds numbers $\Rm\sim10^{29}...10^{31}$. 
The magnetic Prandtl number $\Pm=\Rm/\Re$ for the ICM 
can, therefore, be as large as~$10^{29}$. 
From theory and numerical 
simulations of isotropic MHD turbulence with 
large Prandtl numbers, we know that under these conditions 
a small-scale dynamo operates: magnetic fluctuations are 
amplified by random stretching of the field lines \citep[see a recent account 
of the theory and simulations of small-scale dynamo 
by \citealt{SCTMM_stokes} and references therein]{Jaffe,Roland,Ruzmaikin_Sokoloff_Shukurov}. 
The magnetic energy grows exponentially at the turnover 
rate of the viscous-scale eddies: 
the exponentiation time is $(\lf/\uf)\Re^{-1/2}\sim10^8$~yr. 
The resulting fields have a folded structure: 
they are organized in long thin flux sheets (or ribbons) 
with direction reversals at the resistive scale 
and field lines remaining relatively straight up to the viscous scale 
or, in the nonlinear regime, even to the outer scale of the turbulence. 

This picture is hard to reconcile with the observed 
magnetic fields in clusters. 
The dynamo-generated fields are expected to have reversals 
on subviscous scales down to the resistive scale, which is 
$\lres\sim\Rm^{-1/2}\lf\sim10^4...10^5$~km 
in clusters, a tiny distance. 
Published rotation measure (RM) data for clusters suggest 
tangled fields with $B\sim1...10~\mu$G and a typical reversal scale 
on the order of $1$~kpc \citep{Feretti_etal,Clarke_Kronberg_Boehringer,Taylor_Fabian_Allen}.
Recently, \citet{Vogt_Ensslin} 
used RM measurements from extended radio sources 
in several clusters to extract magnetic-energy spectra. 
The spectra peak at $\sim1$~kpc and decrease at smaller scales, 
so one can reasonably take $1$~kpc to be the 
resolved characteristic scale of field variation. 
While this is about an order of magnitude below the viscous scale, 
it is certainly much larger than the resistive scale.

These apparent inconsistencies between theory and observations  
are a serious problem. The generation of direction-reversing folded fields 
is a fundamental property of random shear. 
It does not depend on the particular character of the 
turbulent flow and does not require very large $\Re$. 
Even if the cluster fields originally owe their existence 
to some external mechanism \citep[e.g.,][]{Kronberg_etal} 
rather than to in situ generation by turbulence, 
any magnetic field introduced 
into the ICM is tangled by turbulence and rendered indistinguishable 
from a small-scale-dynamo--generated field on the 
timescale of~$\sim10^8$~yr. 

The discrepancy between the predictions of MHD models and the 
observed cluster turbulence has prompted us to reexamine  
the MHD approximation and to recognize that it 
is not, in fact, appropriate for the ICM. 
For the cluster plasma, the ion mean free path 
greatly exceeds the resistive scale: 
$\mfp\sim\vth/\nui\sim\Re^{-1/4}\ld\sim1..10$~kpc. 
Collisionless effects are important at scales below $\mfp$,
and the fluid MHD description is not valid. 
This fundamentally alters the dynamics of both velocity and 
magnetic fields at these scales. 

It can be shown that at frequencies below the ion cyclotron 
frequency~$\Omega_i=eB/cm_i$ 
and scales above the ion gyroradius~$\rho_i=\vth/\Omega_i$, 
the equations for the fluid velocity 
$\vu$ and magnetic field $\vB$ have the following general form 
\bea
\label{eq_u}
\rho\,{\diff\vu\over\diff t} &=& -\vdel\(\pperp + {B^2\over8\pi}\) 
+ \vdel\cdot\lt[\vb\vb(\pperp-\ppar)\rt]
+ {\vB\cdot\vdel\vB\over4\pi},\\
\label{eq_B}
{\diff\vB\over\diff t} &=& \vB\cdot\vdel\vu - \vB\vdel\cdot\vu,
\eea
where $\diff/\diff t = \dd/\dd t + \vu\cdot\vdel$, 
$\pperp$ and $\ppar$ are the perpendicular and parallel 
plasma pressures, respectively, and $\vb=\vB/B$. 
We have dropped the diffusion term in \eqref{eq_B}. 
\Eqref{eq_u} is valid provided the ions are magnetized, i.e., 
$\rho_i\ll\mfp$. For the ICM, this requirement 
is satisfied if $B\gg10^{-18}$~G, which is far below 
the observed field strengths of $1...10~\mu$G. 
The lower limit for dynamically important fields is the field strength 
corresponding to the energy of the viscous-scale eddies 
($\sim\rho\vth^2\Re^{-1/2}$), which gives $B\sim10~\mu$G.
Thus, the ICM would be very well magnetized already for 
dynamically weak fields. 

In such a plasma, 
the fundamental property of 
charged particles moving in a magnetic field is the conservation 
of the first adiabatic invariant $\mu=m\vperp^2/2B$.
When $\mfp\gg\rho_i$, 
this conservation is only weakly broken by collisions. 
As long as $\mu$ is conserved, any change 
in $B$ must be accompanied by a proportional change in $\pperp$.
Thus, the emergence of the pressure anisotropy is a natural 
consequence of the changes in the magnetic field strength and vice 
versa. The following heuristic argument reveals the connection. 
Summing up the first adiabatic invariants of all particles, 
we get $\pperp/B=\const$. Then 
\bea
{1\over\pperp}{\diff\pperp\over\diff t} 
\sim {1\over B}{\diff B\over\diff t} 
-\nui\,{\pperp-\ppar\over p},
\label{dlnB_approx}
\eea
where we have assumed that $\rho=\const$ and 
the pressure anisotropy $\pperp-\ppar$ 
is relaxed by collisions at the rate $\nui$ and remains 
small compared with the total pressure $p=\rho\vth^2$. 
From \eqref{eq_B}, the field strength evolves according~to
\bea
\label{eq_lnB}
{1\over B}{\diff B\over\diff t} = \vb\vb:\vdel\vu 
\eea
(assuming incompressibility $\vdel\cdot\vu=0$ in accordance 
with the earlier observation that the inertial-range motions 
are subsonic).
Using this and \eqref{dlnB_approx}, we get 
$\pperp-\ppar\sim(\rho\vth^2/\nui)\vb\vb:\vdel\vu$. 

A more formal kinetic calculation of the pressure anisotropy can be 
done starting from the kinetic equation for the magnetized ions 
\citep[to the lowest order in $k\rho_i$; see][]{Kulsrud_HPP} 
\bea
\nonumber
{\diff f\over\diff t} &+& \vpar\delpar f + 
{\vperp\over2}\(\vb\vb:\vdel\vu - \vdel\cdot\vu + \vpar{\delpar B\over B}\)
{\dd f\over\dd\vperp}\\
&-&\lt[\vb\cdot\({\diff\vu\over\diff t}+\vpar\delpar\vu\) 
+ {\vperp^2\over2}{\delpar B\over B}\rt]{\dd f\over\dd\vpar} 
= C[f,f],
\label{kin_eq}
\eea
where $\delpar=\vb\cdot\vdel$, $\vperp$ and $\vpar$ are velocities 
perpendicular and parallel to $\vB$ and the right-hand side is the collision term. 
A perturbation theory is constructed under the assumption that the fluid velocity 
is much smaller than the thermal speed and varies at scales 
much longer than the mean free path. This is appropriate for  
the viscous range of Kolmogorov turbulence in cluster-type plasmas. 
Assuming Kolmogorov scaling, the velocity at the viscous scale is $u\sim\Re^{-1/4}\vth$. 
The wave number associated with the viscous scale is 
$\kd\sim\Re^{3/4}\lf^{-1}\sim\Re^{-1/4}\mfp^{-1}$.
We introduce the small parameter 
$\epsilon \sim \kd\mfp \sim u/\vth \sim \Re^{-1/4}$.
In cluster plasmas, $\epsilon\sim0.1...0.3$ ---
while this is not really very small, it is convenient 
to order all terms in \eqref{kin_eq} with respect to $\epsilon$, 
which we believe to be the essential physical small parameter 
in the problem. Namely, we assume 
$\kpar\mfp\sim\epsilon$ and 
$d/dt\sim\nabla u\sim\epsilon^2\vth/\mfp$ 
(this orders out compressible motions, so $\vdel\cdot\vu=0$)
and take the zeroth-order distribution function $f_0$ to be a Maxwellian. 
For simplicity, we use the Lorentz pitch-angle-scattering 
form of the linearized collision operator 
$C[f_0,f]=(1/2)\nui(\vth/v)^3(\dd/\dd\xi)(1-\xi^2)\dd f/\dd\xi$,
where $\xi=\vpar/v$ and the derivatives are at constant 
$v=(\vperp^2+\vpar^2)^{1/2}$ \citep[e.g.,][]{Helander_Sigmar}. 
Then the anisotropic ion distribution function up to order 
$\epsilon^2$ is
\bea
f(\vperp,\vpar)={n\,\exp(-v^2/2\vth^2)\over(2\pi\vth^2)^{3/2}}
\lt[1+{\vb\vb:\vdel\vu\over\nui}
{v^3\bl(v^2-3\vpar^2\br)\over3\vth^5}\rt].
\label{f_sln}
\eea
Computing $\pperp$ and $\ppar$, we obtain 
the result that is the lowest order (in $k\rho_i$) term in 
the plasma pressure tensor first derived by \citet{Braginskii} 
(who used a somewhat different ordering in his perturbation scheme)
\bea
\pperp-\ppar = 3\rho\nuB\vb\vb:\vdel\vu \equiv \rho\vth^2\Delta, 
\label{dp_brag}
\eea
where $\Delta\sim\epsilon^2$ is the dimensionless measure of the 
pressure anisotropy and 
$\nuB$ is the Braginskii ion viscosity, 
which is analogous to the ordinary viscosity for the unmagnetized case: 
$\nuB\sim\vth\mfp\sim\vth^2/\nui$. It is this viscosity that we 
used in our definition of $\Re$ and the viscous scale. 
For simplicity, $T_e=T_i\equiv T$, so the electron contribution 
to the pressure anisotropy is subdominant by a factor 
of $(m_e/m_i)^{1/2}\simeq0.02$. 

\Eqsref{eq_u}, \exref{eq_B}, and \exref{dp_brag}, together with 
$\vdel\cdot\vu=0$, are 
a closed set. The Braginskii viscosity has two 
key properties. First, the velocities that it 
dissipates are those that change the strength of the magnetic 
field [\eqref{eq_lnB}], so the small-scale dynamo action 
is associated only with velocities above the viscous scale. 
Second, velocity gradients transverse to the magnetic field 
(e.g., shear-Alfv\'en--polarized fluctuations) are undamped. 
Velocity fluctuations can now penetrate below the viscous cutoff. 
In what follows, we will show that the Braginskii 
viscosity not only fails to damp all kinetic energy 
at the viscous scale but triggers fast-growing instabilities at 
subviscous scales. 

It has, in fact, been known for a long time that anisotropic 
distributions lead to instabilities \citep{Rosenbluth,Chandrasekhar_Kaufman_Watson,Parker1,Vedenov_Sagdeev}. 
A vast literature exists on these instabilities and their 
(mostly space physics) applications, 
which we do not attempt to review. The standard approach 
is to postulate a bi-Maxwellian equilibrium 
distribution with $\Tperp\neq\Tpar$ 
\cite[e.g.,][and references therein]{Gary_book,Ferriere_Andre}. 
We do not need to adopt such a description because 
\eqsref{eq_u}, \exref{eq_B}, and \exref{dp_brag} 
incorporate the pressure anisotropy in a self-consistent way. 

Let us imagine a turbulent cascade that originates from 
the large-scale driving, extends down to the viscous scale, and 
gives rise to velocity and magnetic fields $\vu$ and $\vB$ that vary 
on time scales $\gtrsim(\nabla u)^{-1}$ and on spatial 
scales $\gtrsim\kd^{-1}$. 
We study the stability of such fields. 
The presence of turbulent shear (velocity gradients)
gives rise to the pressure anisotropy given by \eqref{dp_brag}. 
We now look for linear perturbations $\dvu$, $\dvB$, 
$\dpperp$, and $\dppar$ that have frequencies $\omega\gg\nabla u$
and wavenumbers $k\gg\kd$. 
With respect to these perturbations, the unperturbed 
rate-of-strain tensor $\vdel\vu$ can be viewed as constant in space 
and time.\footnote{\citet{Balbus} recently considered 
an instability triggered by Braginskii viscosity when
the unperturbed $\vdel\vu$ consists of rotation and radial shear
in a Keplerian disk \citep[see also][]{Sharma_Hammett_Quataert}. 
His instability had a growth rate $\gamma\sim|\vdel\vu|$. 
In contrast, we are interested in instabilities that are 
much faster than the rate of strain 
and arise when turbulent stretching is present.} 
Linearizing \eqref{eq_u} and neglecting temporal and 
spatial derivatives of the unperturbed quantities, we get
\bea
\nonumber
-i\omega\rho\dvu &=& -i\vk\,\delta\(\pperp+{B^2\over8\pi}\) + 
\(\pperp-\ppar+{B^2\over4\pi}\)\delta\(\vb\cdot\vdel\vb\)\\ 
&&+\,\,\vb\,\delta\lt[\delpar\(\pperp-\ppar\)
-\(\pperp-\ppar - {B^2\over4\pi}\){\delpar B\over B}\rt],\qquad
\label{eq_u_lin} 
\eea
where $\delta\(\vb\cdot\vdel\vb\)=i\kpar\dvb$ 
and from the linearized \eqref{eq_B},
$\dvb=-(\kpar/\omega)\dvuperp$ and 
$\dB/B=\vkperp\cdot\dvuperp/\omega$. 
In the resulting dispersion relation, it is always 
possible to split off the part that corresponds to 
the modes that have shear-Alfv\'en--wave polarization, 
$\dvu\propto\vkperp\times\vb$, 
\bea
\label{dr_shear}
\omega^2 = \kpar^2\vth^2\(\Delta + 2\beta^{-1}\),
\eea
where $\beta=2\vth^2/\vA^2=8\pi\rho\vth^2/B^2$.
When the magnetic energy is larger than the energy of the 
viscous-scale eddies, $\beta\ll|\Delta|^{-1}\sim\Re^{1/2}$, 
\eqref{dr_shear} describes shear Alfv\'en waves. 
In the opposite limit, $\beta\gg|\Delta|^{-1}$, 
and if $\Delta<0$, an instability appears with the growth rate 
$\gamma=\kpar\vth|\Delta|^{1/2}\sim\epsilon\kpar\vth$. 
Since $\kpar\gg\kd$, the instability is faster than the rate of 
strain, $\gamma\gg\nabla u\sim\epsilon\kd\vth$, in 
accordance with the assumption made in the derivation of \eqref{dr_shear}. 
This instability, triggered by $\ppar>\pperp$, is called the firehose instability. 
It does not depend on the way the pressure perturbations are calculated 
because it arises from the perturbation of the field-line curvature 
$\vb\cdot\vdel\vb$ in \eqref{eq_u_lin}: to linear order, 
it entails no perturbation of the field strength 
and therefore does not alter the pressure. 

In order to determine the stability of perturbations with other polarizations,
$\dpperp$ and $\dppar$ have to be computed. We retain the assumption that 
$\omega\gg\nabla u$ and consider scales smaller than the mean free path, $k\mfp\gg1$. 
At these scales a fully kinetic description must be used. 
It is sufficient to look for ``subsonic'' perturbations such 
that $\omega\ll k\vth$ because the high-frequency perturbations 
are subject to strong collisionless damping and cannot 
be rendered unstable by a small anisotropy. 
We linearize \eqref{kin_eq} around the distribution function~\exref{f_sln} 
and calculate $\dpperp$ and $\dppar$ from the perturbed distribution 
function while neglecting terms of order $(\omega/\kpar\vth)^3$ and higher. 
This leads~to
\bea
\label{dr_kin}
\omega^2\dvu &=& \kpar^2\vth^2\(\Delta+2\beta^{-1}\)\dvuperp
-\vb\,{\omega^2\over\kpar}\,\vkperp\cdot\dvuperp\\ 
\nonumber
&&+\ \vkperp\lt[2\(-\Delta + \beta^{-1}\)\vth^2 
- i\omega\sqrt{2\pi}\,{\vth\over|\kpar|}
+ {2\,{\omega^2\over\kpar^2}}\rt]\vkperp\cdot\dvuperp.
\nonumber
\eea 
For the modes with shear-Alfv\'en--wave polarization, we recover \eqref{dr_shear}.
If we now dot \eqref{dr_kin} with $\vb$, we get 
$\dupar=-\vkperp\cdot\dvuperp/\kpar$. Therefore, 
the perturbations with $\dupar\neq0$ are incompressible 
and slow-wave polarized. 
For these perturbations, $\dvuperp\propto\vkperp$, 
and the dispersion relation is 
\bea
\nonumber
\biggl(1-{2\kperp^2\over\kpar^2}\biggr)\omega^2 
&+& i\omega\sqrt{2\pi}\,{\kperp^2\vth\over|\kpar|}\\ 
&-& \lt[\(\kpar^2-2\kperp^2\)\Delta + 2k^2\beta^{-1}\rt]\vth^2 = 0.
\label{dr_slow_kin}
\eea
In the weak-field limit ($\beta\gg|\Delta|^{-1}$), 
the third term leads to instability. If $\kperp^2/\kpar^2\ll|\Delta|^{1/2}$ 
(the case of parallel or nearly parallel propagation), 
the second term is negligible 
and, for $\Delta<0$, we get $\gamma\simeq\kpar\vth|\Delta|^{1/2}$, 
as in the case of the firehose instability. 

When $\kperp$ is not small (oblique propagation),
the second term in \eqref{dr_slow_kin} dominates the first. 
It contains the effect of the collisionless kinetic damping due to 
resonant wave-particle interactions \citep{Barnes} 
(a physical discussion of the role of resonant and nonresonant 
particles in the mirror instability can be found in \citealt{Southwood_Kivelson}).
The growth rate~is
\bea
\label{gamma_mirror}
\gamma = \(2\over\pi\)^{1/2}|\kpar|\vth\lt[ 
\Delta\(1-{\kpar^2\over2\kperp^2}\) 
- \beta^{-1}\(1+{\kpar^2\over\kperp^2}\)\rt].
\eea 
Modes with $\kpar>\sqrt{2}\kperp$ are unstable if 
$\Delta<0$ (slow-wave--polarized firehose instability); 
modes with $\kperp>\kpar/\sqrt{2}$ are unstable if 
$\Delta>0$ (mirror instability). The growth rate in either case is 
$\gamma\sim \kpar\vth|\Delta|\sim\epsilon^2 \kpar\vth$. 

The instabilities described above have growth rates 
proportional to $\kpar$. Thus, MHD equations with Braginskii 
viscosity do not constitute a well-posed problem. This means, 
for example, that any numerical simulation of these equations 
will blow up at the grid scale unless a small additional 
isotropic viscosity is introduced. 
The linear-in-$\kpar$ behavior of the growth rate is only 
modified if the finiteness of the gyroradius $\rho_i$ is 
taken into account. This is done using the 
general plasma dispersion relation \citep[e.g.,][]{Davidson_HPP}
with the equilibrium distribution function~\exref{f_sln}.
The growth rate of slow-wave--polarized instabilities 
[\eqref{gamma_mirror}] peaks at $k\rho_i\sim1$ with
$\gmax\sim\Delta\Omega_i$. 
For the shear-Alfv\'en--polarized firehose, 
the fastest growth is for $\kperp/\kpar\simeq\sqrt{2/3}$, 
and the peak growth rate $\gmax\sim|\Delta|^{1/2}\Omega_i$ is achieved 
at $k\rho_i\sim1$. We omit the derivation of these results. 

We have shown that, given sufficiently high $\beta$, 
firehose and mirror instabilities occur in the regions 
of decreasing ($\Delta<0$) and increasing ($\Delta>0$) 
magnetic field strength, respectively. 
The viscous-scale motions associated with the turbulent 
cascade from large scales will stretch the magnetic field 
and thereby produce regions of both types: the typical field 
structure resulting from random stretching 
is folded flux sheets (or ribbons) with field amplification 
regions containing relatively straight 
direction-alternating fields and curved corners 
where the field is weaker \citep[the field curvature 
and strength are anticorrelated; see][]{SCTMM_stokes}. 
If $B\gtrsim10^{-18}$~G, the plasma is magnetized ($\rho_i<\mfp$) 
and this structure is intrinsically unstable: straight growing fields to the mirror, 
curved weakening fields to the firehose instability. 
Since the instabilities are much faster than the turbulent stretching, 
their growth and saturation will have a profound effect on the structure of 
the field. We do not yet have a detailed theory of this 
process and only give a qualitative discussion of what can plausibly 
happen. 

When turbulent stretching acts on some initial 
weak field $B_0$, pressure anisotropies arise, 
$\Delta_0\sim(1/\nui)d\ln B_0/dt$ [see \eqref{dlnB_approx}],
and drive mirror and firehose instabilities. 
Both produce fluctuating fields $\dvB$ curved at scale 
$\rho_{i,0}\sim3\times10^3(10^{-18}~{\rm G}/B_0)$~pc. 
Let us concentrate on the mirror 
instability as it amplifies field strength in the linear order. 
The amplification rate is 
$\gamma_0\sim\Delta_0\Omega_{i,0}\sim10^{-8}(B_0/10^{-18}~{\rm G})~{\rm yr}^{-1}$. 
Suppose for a moment that the instability does not saturate 
until the field strength increases to some value $B_1\sim B_0+\dB$,
where $\dB/B_0\sim1$. Then 
the field at scale $\rho_{i,0}$ is unstable to perturbations 
at scale $\rho_{i,1}\sim (B_0/B_1)\rho_{i,0}<\rho_{i,0}$. 
This is a secondary mirror instability that feeds on the 
pressure anisotropy due to the increasing field $B_1$ excited by the primary. 
Braginskii theory cannot be used to calculate this anisotropy 
because the field $B_1$ is curved on a collisionless scale $\rho_{i,0}$. 
However, $\mu$ is still conserved at this scale 
(conservation 
is broken at scale $\rho_{i,1}$) and we can adapt the estimate~\exref{dlnB_approx} 
to the collisionless regime by assuming that pressure anisotropy is 
relaxed in the time particles streaming along the field line 
need to cover the distance $\rho_{i,0}$: we replace $\nui$
in \eqref{dlnB_approx} by $\vth/\rho_{i,0}=\Omega_{i,0}$ and write 
$\Delta_1\sim(1/\Omega_{i,0})d\ln B_1/dt\sim\gamma_0/\Omega_{i,0}\sim\Delta_0$. 
Thus, the pressure anisotropies driving the secondary and the primary instabilities  
are the same. The growth rate of the perturbations at scale $\rho_{i,1}$ 
is then $\gamma_1\sim\Delta_0\Omega_{i,1}\sim (B_1/B_0)\gamma_0$. 
This argument can be iterated. As the growth rate $\gmax\propto B$, 
magnetic field grows explosively, $dB/dt\propto B^2$, until it is strong enough 
to cancel the pressure anisotropy and shut down the instabilities: 
$\beta\sim\Delta^{-1}\sim\Re^{1/2}$. This gives 
$B\sim10~\mu$G, which is the lower bound for dynamically 
important fields and is comparable to the observed field 
strength in clusters. Thus, a seed field of order $10^{-18}$~G 
\citep[attainable by primordial mechanisms; 
see, e.g.,][]{Gnedin_Ferrara_Zweibel} can be amplified 
to dynamical strength in about $10^8$~yr.  

The above argument depends on the assumption that fluctuating 
fields would grow to $\dB/B_0\sim1$. What if 
the saturation of the instabilities is quasilinear, with 
$|\dvB|^2/B_0^2\ll1$? 
In the quasilinear theory, small fluctuations scatter particles and effectively 
increase their collision frequency to $\nuQL\sim\(|\dvB|^2/B_0^2\)\gmax$. 
The pressure is isotropized and the instabilities are quenched. 
Since the effective viscosity ($\sim\vth^2/\nuQL$) of the plasma is reduced, 
the Reynolds number increases $\ReQL\sim\(\lf/\vth\)\nuQL\propto B$ 
and the turbulent cascade extends to smaller scales, 
giving rise to faster field stretching 
(by the viscous-scale eddies at the rate $\propto\ReQL^{1/2}$) 
and to pressure anisotropy $\Delta\sim\ReQL^{-1/2}$. 
This argument can be formalized somewhat and gives explosive growth 
of both $\ReQL$ and $B$ until $\beta\sim\ReQL^{1/2}$. At this point 
the instabilities should start shutting down with $\Re$ dropping 
back to its original value based on particle collisions and magnetic field 
following up so that $\beta\sim\Re^{1/2}$. The characteristic time 
for this process to complete itself is $\sim10^8$~yr. 
Thus, both the quasilinear scenario ($\dB/B_0\ll1$) and the case of $\dB/B_0\sim1$ 
appear to produce a fast amplification stage with similar end results. 

After this stage, $\beta\lesssim\Re^{1/2}$, 
the field can be stretched without going unstable
and the folded structure could finally be set up. 
However, differences remain between the MHD model with isotropic viscosity 
\citep{SCTMM_stokes} and the ICM. 
First, the Braginskii viscosity implies that a cascade of 
shear Alfv\'en waves 
may exist between the viscous scale and the ion gyroscale. 
The properties of this cascade remain to be understood. 
Second, even when the rms field strength $\Brms$ is sufficient 
to quench the instabilities, there will always be regions 
where the field is weak. For the folded fields, these are the 
bending regions (curved field). They can become unstable unless the 
field-line curvature there is larger than $1/\rbend$, where 
$\rbend$ is the gyroradius computed with the local field strength 
$\Bbend\ll\Brms$ and, therefore, is larger than $\rho_i$ based on $\Brms$
by a factor of $\Brms/\Bbend$. 
Based on the idea that all characteristic scales are fixed by 
equating the curvature in the bending regions to $1/\rbend$, 
we have constructed a field-reversal--scale estimate of order 
$10$~pc \citep{SCKHS_cracow}. 
However, our argument assumed that the field 
structure is exactly the same as in the case of isotropic viscosity 
with just the magnetic cutoff scale undetermined. In order to justify 
this assumption, or to learn otherwise, we will need a more 
quantitative theory. 

Thus, while it is not hard to envision how magnetic fields in clusters 
can reach observed strengths on a fairly short timescale, 
the full understanding of field structure requires more work.

\acknowledgements

We would like to thank E.~Quataert for useful discussions.
This work was supported by 
a UKAFF Fellowship (A.A.S.), 
the Department of Energy (DOE) contract DE-AC02-76CH03073 (G.W.H.\ and P.S.), 
the NSF grant AST~00-98670, 
and the DOE Center for Multiscale Plasma Dynamics.


\begin{thebibliography}{}

\bibitem[Balbus(2004)]{Balbus}
Balbus, S.~A. 2004, \apj, 616, 857

\bibitem[Barnes(1966)]{Barnes}
Barnes, A. 1966, Phys.\ Fluids, 9, 1483

\bibitem[Braginskii(1965)]{Braginskii}
Braginskii, S.~I. 1965, Rev.\ Plasma Phys., 1, 205 


\bibitem[Chandrasekhar et al.(1958)]{Chandrasekhar_Kaufman_Watson}
Chandrasekhar, S., Kaufman, A.~N., \& Watson, K.~M. 1958, 
Proc.\ R.~Soc.\ London Ser.~A, 245, 435

\bibitem[Clarke et al.(2001)]{Clarke_Kronberg_Boehringer}
Clarke, T.~E., Kronberg, P.~P., \& B\"ohringer, H. 2001, 
\apjl, 547, L111

\bibitem[Davidson(1983)]{Davidson_HPP}
Davidson, R.~C. 1983, 
in Handbook of Plasma Physics, Vol.~1, 
ed.\ A.~A.~Galeev \& R.~N.~Sudan 
(Amsterdam: North--Holland),~519 




\bibitem[Feretti et al.(1995)]{Feretti_etal}
Feretti, L., Dallacasa, D., Giovannini, G., \& Tagliani, A. 1995, 
\aap, 302, 680 

\bibitem[Ferri\`ere \& Andr\'e(2002)]{Ferriere_Andre}
Ferri\`ere, K.~M. \& Andr\'e, N. 2002, 
J.~Geophys.\ Res., 107, 1349

\bibitem[Gary(1993)]{Gary_book}
Gary, S.~P. 1993, Theory of Space Plasma Microinstabilities 
(Cambridge: Cambridge Univ.\ Press)

\bibitem[Gnedin et al.(2000)]{Gnedin_Ferrara_Zweibel}
Gnedin, N.~Y., Ferrara, A., \& Zweibel, E. 2000, \apj, 539, 505


\bibitem[Helander \& Sigmar(2002)]{Helander_Sigmar}
Helander, P. \& Sigmar, D.~J. 2002, 
Collisional Transport in Magnetized Plasmas 
(Cambridge: Cambridge Univ.\ Press)

\bibitem[Inogamov \& Sunyaev(2003)]{Inogamov_Sunyaev}
Inogamov, N.~A. \& Sunyaev, R.~A. 2003, 
Astron.\ Lett., 29, 791

\bibitem[Jaffe(1980)]{Jaffe}
Jaffe, W. 1980, \apj, 241, 925


\bibitem[Kronberg et al.(2001)]{Kronberg_etal}
Kronberg, P.~P., Dufton, Q.~W., Li, H., \& Colgate, S.~A. 2001,
\apj, 560, 178

\bibitem[Kulsrud(1983)]{Kulsrud_HPP}
Kulsrud, R.~M. 1983, 
in Handbook of Plasma Physics, Vol.~1, 
ed.\ A.~A.~Galeev \& R.~N.~Sudan 
(Amsterdam: North--Holland),~115 



\bibitem[Norman \& Bryan(1999)]{Norman_Bryan}
Norman, M.~L. \& Bryan, G.~L. 1999, Lecture Notes in Physics 530, 
The Radio Galaxy Messier 87, ed.\ H.-J.\ R\"oser \& 
K.\ Meisenheimer (New York: Springer), 106

\bibitem[Parker(1958)]{Parker1}
Parker, E.~N. 1958, Phys.\ Rev., 109, 1874



\bibitem[Ricker \& Sarazin(2001)]{Ricker_Sarazin}
Ricker, P.~M. \& Sarazin, C.~L. 2001, \apj, 561, 621

\bibitem[Roettinger et al.(1999)]{Roettinger_Stone_Burns}
Roettinger, K., Stone, J.~M., \& Burns, J.~O. 1999, 
\apj, 518, 594

\bibitem[Roland(1981)]{Roland}
Roland, J. 1981, \aap, 93, 407

\bibitem[Rosenbluth(1956)]{Rosenbluth}
Rosenbluth, M.~N. 1956, Los Alamos Laboratory Report LA-2030 


\bibitem[Ruzmaikin et al.(1989)]{Ruzmaikin_Sokoloff_Shukurov}
Ruzmaikin, A., Sokoloff, D., \& Shukurov, A. 1989, 
\mnras, 241, 1



\bibitem[Schekochihin et al.(2004)]{SCTMM_stokes}
Schekochihin, A.~A., Cowley, S.~C., Taylor, S.~F., 
Maron, J.~L., \& McWilliams, J.~C. 2004, 
\apj, 612, 276 

\bibitem[Schekochihin et al.(2005)]{SCKHS_cracow}
Schekochihin, A.~A., Cowley, S.~C., Kulsrud, R.~M., Hammett, G.~W., 
and Sharma, P. 2005, 
in The Magnetized Plasma in Galaxy Evolution, 
ed.\ K.~T.~Chyzy et al. 
(Cracow: Jagiellonian Univ.), 86 (astro-ph/0411781)

\bibitem[Schuecker et al.(2004)]{Schuecker_etal}
Schuecker, P., Finoguenov, A., Miniati, F., B\"ohringer, H., 
\& Briel, U.~G. 2004, \aap, 426, 387


\bibitem[Sharma et al.(2003)]{Sharma_Hammett_Quataert}
Sharma, P., Hammett, G.~W., \& Quataert, E. 2003, \apj, 596, 1121 

\bibitem[Southwood \& Kivelson(1993)]{Southwood_Kivelson}
Southwood, D.~J. \& Kivelson, M.~G. 1993, 
J.~Geophys.\ Res., 98, 9181

\bibitem[Spitzer(1962)]{Spitzer_book}
Spitzer, L. 1962, Physics of Fully Ionized Gases 
(New York: Wiley) 


\bibitem[Taylor et al.(2002)]{Taylor_Fabian_Allen}
Taylor, G.~B., Fabian, A.~C., \& Allen, S.~W. 2002, MNRAS, 334, 769

\bibitem[Vedenov \& Sagdeev (1958)]{Vedenov_Sagdeev}
Vedenov, A.~A. \& Sagdeev, R.~Z. 1958, 
Soviet Phys.---Dokl., 3, 278

\bibitem[Vogt \& En{\ss}lin(2003)]{Vogt_Ensslin}
Vogt, C. \& En{\ss}lin, T.~A. 2003, \aap, 412, 373


\end{thebibliography}
\end{document}